\documentclass [twocolumn]{aastex6}
\usepackage[encapsulated]{CJK}
\usepackage{setspace}
\usepackage{array}
\newcolumntype{C}[1]{>{\centering}p{#1}}

\begin{document}
\begin{CJK*}{GBK}{song}
\renewcommand{\thefootnote}{\fnsymbol{footnote}}


\title{The {$^{95}$Z\MakeLowercase{r}($\MakeLowercase{n}$, $\gamma$)$^{96}$Z\MakeLowercase{r}} cross section from the surrogate ratio method and its effect on the s-process nucleosynthesis}


\author{S. Q. Yan\altaffilmark{1}, Z. H. Li\altaffilmark{1}, Y. B. Wang\altaffilmark{1}, K. Nishio\altaffilmark{2}, M. Lugaro\altaffilmark{3,4}, A. I. Karakas\altaffilmark{4}, H. Makii\altaffilmark{2}, P. Mohr\altaffilmark{5,6}, J. Su\altaffilmark{1}, Y. J. Li\altaffilmark{1}, I. Nishinaka\altaffilmark{2}, K. Hirose\altaffilmark{2}, Y. L. Han\altaffilmark{1}, R. Orlandi\altaffilmark{2}, Y. P. Shen\altaffilmark{1}, B. Guo\altaffilmark{1}, S. Zeng\altaffilmark{1}, G. Lian\altaffilmark{1}, Y. S. Chen\altaffilmark{1}, and W. P. Liu\altaffilmark{1}}
\affil{\altaffilmark{1}{China Institute of Atomic Energy, P. O. Box 275(10), Beijing 102413, P. R. China; panyu@ciae.ac.cn}}
\affil{\altaffilmark{2}{Japan Atomic Energy Agency, Tokai, Naka, Ibaraki 319-1195, Japan}}
\affil{\altaffilmark{3}{Konkoly Observatory, Research Centre for Astronomy and Earth Sciences, Hungarian Academy of Sciences, 1121 Budapest, Hungary; maria.lugaro@csfk.mta.hu}}
\affil{\altaffilmark{4}{Monash Centre for Astrophysics, School of Physics and Astronomy, Monash University, Clayton, VIC 3800, Australia}}
\affil{\altaffilmark{5}{Institute for Nuclear Research (ATOMKI), H-4001 Debrecen, Hungary}}
\affil{\altaffilmark{6}{Diakonie-Klinikum, D-74523 Schw\"abisch Hall, Germany}}








\begin{abstract}

The $^{95}$Zr($n$,$\gamma$)$^{96}$Zr reaction cross section is crucial in the modelling of $s$-process nucleosynthesis in asymptotic giant branch stars because it controls the operation of the branching point at the unstable $^{95}$Zr and the subsequent production of $^{96}$Zr. We have carried out the measurement of the $^{94}$Zr($^{18}$O,$^{16}$O) and $^{90}$Zr($^{18}$O,$^{16}$O) reactions and obtained the $\gamma$-decay probability ratio of $^{96}$Zr* and $^{92}$Zr* to determine the $^{95}$Zr($n$,$\gamma$)$^{96}$Zr reaction cross sections with the surrogate ratio method. Our deduced maxwellian-averaged cross section of 66$\pm$16 mb at 30 keV is close to the value recommended by \cite{BAO00}, but 30\% and more than a factor of two larger than the values proposed by \cite{TOU90} and \cite{LUG14}, respectively, and routinely used in $s$-process models. We tested the new rate in stellar models with masses between 2 and 6 M$_\odot$ and metallicities 0.014 and 0.03. The largest changes - up 80\% variations in $^{96}$Zr - are seen in models of mass 3-4 M$_\odot$, where the $^{22}$Ne neutron source is mildly activated. The new rate can still provide a match to data from meteoritic stardust silicon carbide grains, provided the maximum mass of the parent stars is below 4 M$_\odot$, for a metallicity of 0.03.

\end{abstract}




\section{Introduction} \label{sec:intro}

The elements heavier than iron are produced via neutron captures: the \textit{slow} neutron capture process (\textit{s}-process) and the \textit{rapid} neutron capture process (\textit{r}-process), except for a minor contribution from the so-called \textit{p}-process \citep{BUR57,SEE65,WAL97}. The \textit{r}-process is associated with explosive nucleosynthesis in core-collapse supernovae or neutron star mergers. Because of the extremely high neutron densities ($\gg$ $10^{20}$ cm$^{-3}$) the time scale for neutron capture is of the order of milliseconds, and the neutron-capture path involves very neutron rich nuclei, which decay to their corresponding isobars once the neutron flux is extinguished.  The \textit{s}-process is associated with the thermally pulsating asymptotic giant branch (TP-AGB) phases of low-mass ($<$ 8 $M_\odot$) stars \citep{GAL88,MEY94,GAL98,ARL99,BUS99,GOR00,CRI09,BIS11,LUG12,KAR14} and the evolutionary hydrostatic phases of more massive($>$ 10 $M_\odot$) stars \citep{KAP89,RAI91,TRA04,PIG10,FRI12}, which contribute to the main and strong \textit{s}-component from Sr to Pb/Bi and the weak \textit{s}-component between Fe and Sr, respectively. The neutron densities for the \textit{s}-process are of the order of $10^6$-$10^{13}$ cm$^{-3}$, the time scale of neutron capture (order of years) is usually much larger than the average half-lives of $\beta$-unstable nuclei, and the reaction path of \textit{s}-process thus follows the valley of $\beta$ stability.

Zirconium is a typical \textit{s}-process element belonging to the first \textit{s}-process peak and mostly produced by the main component in AGB stars. Its isotopic abundances are sensitive to both the neutron exposure and the neutron density, thus, they are critical to constrain the \textit{s}-process in AGB stars \citep{LUG03}. Because of their near-magic neutron configuration, all isotopes of Zr have relatively small ($n$,$\gamma$) cross sections, therefore, they have comparably high \textit{s} abundances. Zirconium is one of a dozen elements heavier than iron (Kr, Sr, Zr, Mo, Ba, Xe, Nd, Sm, Dy, Eu, W, Pb) whose isotopic abundances can be obtained from the meteoritic  stardust grains with high precision \citep{NIC97,GUN98,BAR07,AVI12a,AVI12b,AVI13}. The Zr abundances can also be spectroscopically observed in cool stars \citep{LAM95} and post-AGB stars \citep{WIN03,SME12}.

The most neutron-rich stable Zr isotope, $^{96}$Zr, is very sensitive to the neutron density during the \textit{s}-process because its production depends on the activation of the branching point at $^{95}$Zr (with a half life of 64 days) for neutron densities above roughly 10$^{10}$ cm$^{-3}$. In low-mass AGB stars, the main neutron source is the $^{13}$C($\alpha$,$n$)$^{16}$O reaction, which produces neutrons with densities lower than 10$^{8}$ cm$^{-3}$. In this condition, the abundance of $^{96}$Zr is depleted by neutron captures and not replenished. On the other hand during the phase of He shell flashes, the temperature can reach up to 3.5 $\times$ $10^8$ K as the stellar mass increases and the $^{22}$Ne neutron source is activated, which results in neutron densities up to 10$^{13}$ cm$^{-3}$ \citep{FIS14}. Since $^{96}$Zr has a low neutron-capture cross section \citep{TAG11b}, once it is produced it tends to accumulate. Thus, the \textit{s}-abundance of $^{96}$Zr can be developed as a tool to probe the neutron density in the stellar interior and interpreted as an indicator for the efficiency of the $^{22}$Ne neutron source and the stellar mass.

To this aim, isotopic abundances obtained from stardust grain and astronomical observation should be compared to the stellar nucleosynthesis calculations. However, while the neutron cross sections of the stable Zr isotopes have been carefully studied \citep{TAG08a,TAG08b,TAG10,TAG11a,TAG11b,TAG13} the cross section of $^{95}$Zr($n$,$\gamma$)$^{96}$Zr reaction is still extremely uncertain. It is hard to measure the cross section directly due to the difficulty of
preparing the target of the short-lived radionuclide $^{95}$Zr. Because of the lack of the experimental data, the theoretical cross section estimates for the Maxwellian Averaged Cross-Section (MACS) at kT = 30 keV vary from 25 mb to 140 mb (KADoNiS v0.3 \footnote[2]{Web site: http://www.kadonis.org.}). The values estimated by \cite{BAO00}, (79 $\pm$ 12 mb at kT = 30 keV), \cite{TOU90}, (50 mb at kT = 30 keV), or \cite{LUG14}(28 mb at kT = 30 keV) are usually adopted in the stellar models. The recommendation by \cite{BAO00} of 79 $\pm$ 12 mb has been adopted in the KADoNiS v0.3 database \citep{DILL09}, but will be replaced in the next version KADoNiS v1.0 \citep{DILL16} by 106 $\pm 26$ mb which is the average from two recent theoretical evaluations TENDL-2015 \citep{TENDL} and ENDF/B-VII.1 \citep{ENDF}. Because of these large discrepancies in the calculation of the MACS of $^{95}$Zr(n,$\gamma$)$^{96}$Zr, improved data are of paramount importance.

Recently, the $^{96}$Zr($\gamma$,$n$)$^{95}$Zr cross section was measured to constrain the $\gamma$-ray strength function, and the HFB-QRPA model was used to obtain the $^{95}$Zr($n$,$\gamma$)$^{96}$Zr cross section \citep{UTS10}. Here, we present the first experimental effort to obtain the ($n$,$\gamma$) cross section of $^{95}$Zr using the surrogate ratio method, based on a benchmark experiment already performed to validate the method \citep{YAN16}.

\section{$^{95}$Z\MakeLowercase{r}(\MakeLowercase{n},$\gamma$)$^{96}$Z\MakeLowercase{r} Cross Sections}
\subsection{Surrogate Ratio Method}
The surrogate ratio method (SRM) is a variation of the surrogate method \citep{YOU03a,YOU03b,PET04,BOY06,KES10}. The SRM has been employed in (n,f) cross sections determination successfully for years \citep{PLE05,BUR06,LYL07,NAY08,LES09,GOL09,RES11}. A comprehensive review can be found in \cite{ESC12} including both the absolute surrogate method and relative ratio method.

Based on the Weisskopf-Ewing limit of the Hauser-Feshbach theory \citep{WEI40}, the $\gamma$-decay probability is independent of the spin-parity of the compound nucleus (CN), therefore, the cross section of $^{95}$Zr($n$,$\gamma$)$^{96}$Zr can be expressed as
  \begin{eqnarray}
  \label{ngcs}
\sigma_{^{95}Zr(n,\gamma)}(E_{n})=\sigma^{CN}_{n+^{95}Zr}(E_{n})G^{CN}_{^{96}Zr^*\rightarrow^{96}Zr+\gamma}(E_{n}).
\end{eqnarray}
In above equation, $\sigma^{CN}_{n+^{95}Zr}(E_{n})$ denotes the CN forming cross section and $G^{CN}_{^{96}Zr^*\rightarrow^{96}Zr+\gamma}(E_{n})$ represents the $\gamma$-decay probability of $^{96}$Zr$^*$, where $E_{n}$ is the incident energy of neutron. Here, $^{96}$Zr$^*$ is formed via a  surrogate reaction: $^{18}$O + $^{94}$Zr $\rightarrow$ $^{16}$O + $^{96}$Zr$^*$, and the $\gamma$ decay of $^{96}$Zr$^*$ is observed in coincidence with the outgoing particle $^{16}$O. The $\gamma$-decay probability of $^{96}$Zr$^*$ can be written as
\begin{eqnarray}
G^{CN}_{^{96}Zr^*\rightarrow^{96}Zr+\gamma}(E_{ex})=\frac{N_{\gamma(^{96}Zr^*)}(E_{ex})}{\epsilon_{\gamma}N_{^{96}Zr^*}(E_{ex})},
\end{eqnarray}
where $N_{^{96}Zr^*}(E_{ex})$ is the total number of $^{96}$Zr$^*$ and $N_{\gamma(^{96}Zr^*)}(E_{ex})$ is the observed number of $^{96}$Zr$^*$ that decays finally into the ground state by emitting $\gamma$-rays. The $E_{ex}$ symbol denotes the excitation energy of $^{96}$Zr$^*$ and $\epsilon_{\gamma}$ is the efficiency of the $\gamma$ detector.

To derive the $^{95}$Zr($n$,$\gamma$)$^{96}$Zr reaction cross section using the SRM, a reference reaction with known cross section is needed. We chose $^{91}$Zr($n$,$\gamma$)$^{92}$Zr as the reference reaction, then the ratio of the two reaction cross sections is
\begin{eqnarray}
\label{ngratio}
&&\frac{\sigma_{^{95}Zr(n,\gamma)}(E_{n})}{\sigma_{^{91}Zr(n,\gamma)}(E_{n})}=\frac{\sigma^{CN}_{n+^{95}Zr}(E_{n})\times G^{CN}_{^{96}Zr^*\rightarrow^{96}Zr+\gamma}(E_{n})}{\sigma^{CN}_{n+^{91}Zr}(E_{n})\times G^{CN}_{^{92}Zr^*\rightarrow^{92}Zr+\gamma}(E_{n})}\nonumber\\&&\approx\frac{G^{CN}_{^{96}Zr^*\rightarrow^{96}Zr+\gamma}(E_{n})}
{G^{CN}_{^{92}Zr^*\rightarrow^{92}Zr+\gamma}(E_{n})}\nonumber\\&&=\frac{\epsilon_{\gamma(^{92}Zr^*)}N_{\gamma(^{96}Zr^*)}(E_{n})}
{\epsilon_{\gamma(^{96}Zr^*)}N_{\gamma(^{92}Zr^*)}(E_{n})}\times\frac{N_{^{92}Zr^*}(E_{n})}{N_{^{96}Zr^*}(E_{n})}.
\end{eqnarray}
The reference reaction is chosen similar to the desired reaction so that their CN formation cross sections are almost same. A theoretical calculation using the UNF code\citep{ZHA92,ZHA93,ZHA02} shows that the two CN formation cross sections are almost equal within 7$\%$ difference. Thus, $\sigma^{CN}_{n+^{95}Zr}(E_{n})/\sigma^{CN}_{n+^{91}Zr}(E_{n}) \approx 1$ and the ($n$, $\gamma$) cross section ratio can be simplified to the ratio of $\gamma$-decay probabilities.

In the SRM experiment we use $^{94}$Zr($^{18}$O, $^{16}$O)$^{96}$Zr$^*$ and $^{90}$Zr($^{18}$O, $^{16}$O)$^{92}$Zr$^*$ reactions to form the compound nuclei $^{96}$Zr$^*$ and $^{92}$Zr$^*$, respectively. The ratio $N_{^{92}Zr^*}(E_{ex})/N_{^{96}Zr^*}(E_{ex})$ in Eq. \ref{ngratio} can be determined from the CN formation cross section integrated over the detector solid angle $\sigma^{CN}_{Zr^*}(E_{ex})$, the thickness of target $\rho$, the beam current $I$ and the efficiency of the particle detector $\epsilon_{p}$ through the relation
\begin{eqnarray}
\label{cnratio}
&&\frac{N_{^{92}Zr^*}(E_{ex})}{N_{^{96}Zr^*}(E_{ex})}=\frac{\sigma^{CN}_{^{90}Zr(^{18}O, ^{16}O)^{92}Zr^*}(E_{ex})}
{\sigma^{CN}_{^{94}Zr(^{18}O, ^{16}O)^{96}Zr^*}(E_{ex})}\nonumber\\&&\times \frac{\rho_{(^{90}Zr)}\times I_{^{18}O(^{90}Zr)}\times\epsilon_{p(^{90}Zr)}}{\rho_{(^{94}Zr)}\times I_{^{18}O(^{94}Zr)}\times\epsilon_{p(^{94}Zr)}}.
\end{eqnarray}
We identified the ejectile nucleus $^{16}$O with a silicon telescope to determine the excitation energy of corresponding CN, $\epsilon_{p(^{90}Zr)}$ and $\epsilon_{p(^{94}Zr)}$ can be canceled since the two surrogate reactions are measured in the same experimental set-up. The two surrogate reactions are chosen to be similar so that $\sigma^{CN}_{^{92}Zr^*}(E_{ex})/\sigma^{CN}_{^{96}Zr^*}(E_{ex}) \approx 1$,  Eq. \ref{ngratio} becomes
\begin{eqnarray}
\label{csratio}
&&\frac{\sigma_{^{95}Zr(n,\gamma)}(E_{n})}{\sigma_{^{91}Zr(n,\gamma)}(E_{n})}\approx\nonumber\\&&\frac{N_{\gamma(^{96}Zr^*)}(E_{n})}{N_{\gamma(^{92}Zr^*)}(E_{n})}\times
\frac{\epsilon_{\gamma(^{92}Zr^*)}\times\rho_{(^{90}Zr)}\times I_{^{18}O(^{90}Zr)}}{\epsilon_{\gamma(^{96}Zr^*)}\times\rho_{(^{94}Zr)}\times I_{^{18}O(^{94}Zr)}}\nonumber\\
&&\approx C_{nor}\frac{N_{\gamma(^{96}Zr^*)}(E_{n})}{N_{\gamma(^{92}Zr^*)}(E_{n})}.
\end{eqnarray}
The  normalization factor $C_{nor}$ can be evaluated by correcting the target thickness, the beam current and the $\gamma$-ray efficiency $\epsilon_{\gamma(^{92}Zr^*)}$, $\epsilon_{\gamma(^{96}Zr^*)}$ of the two surrogate reactions. After $N_{\gamma(^{96}Zr^*)}(E_{n})$ and $N_{\gamma(^{92}Zr^*)}(E_{n})$ are obtained experimentally, the cross section of the $^{95}$Zr($n$,$\gamma$)$^{96}$Zr reaction can be extracted using the known cross section of the reference reaction $^{91}$Zr($n$,$\gamma$)$^{92}$Zr. Since the total number of the compound nucleus ${N_{Zr^*}(E_{ex})}$ is not needed in SRM, uncertainties arising from ${N_{Zr^*}(E_{n})}$ are eliminated.

\subsection{Benchmark Experiment}

\begin{figure}[ht!]
\figurenum{1}
\label{93ZrCS}
\plotone{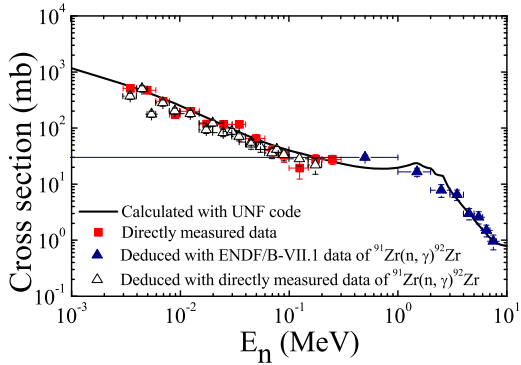}
\caption{Cross section of $^{93}$Zr($n$,$\gamma$)$^{94}$Zr. The red squares are directly measured data \citep{MAC85}, the open triangles are deduced by the $\gamma$-decay probability ratio and $^{91}$Zr($n$,$\gamma$)$^{92}$Zr cross sections \citep{MUS77}, the blue triangles are deduced by the $\gamma$-decay probability ratios and the $^{91}$Zr($n$,$\gamma$)$^{92}$Zr ENDF/B-VII.1 cross section, solid curve is the results calculated by UNF code. }
\end{figure}

Before this work, a benchmark experiment \citep{YAN16} was carried out to check the validity of SRM in (n, $\gamma$) cross section determinations. In that experiment, the $^{91}$Zr($n$,$\gamma$)$^{92}$Zr was chosen as the reference reaction and the $^{90}$Zr($^{18}$O,$^{16}$O) and $^{92}$Zr($^{18}$O,$^{16}$O) reactions were taken as the surrogate reactions to populate the neutron resonance states in $^{92}$Zr and $^{94}$Zr in a large range of equivalent neutron energies. The relative $\gamma$-decay probability ratios of the neutron resonance states in $^{94}$Zr and $^{92}$Zr were measured, and the cross section of the $^{93}$Zr($n$,$\gamma$)$^{94}$Zr reaction was derived from the experimentally obtained ratios and the cross sections of the $^{91}$Zr($n$,$\gamma$)$^{92}$Zr reaction in the equivalent neutron energy range of $E_n$ = 0 - 8 MeV. The deduced cross sections of $^{93}$Zr($n$,$\gamma$)$^{94}$Zr reaction agree with the directly measured ones in the low-energy region, as shown in Fig. \ref{93ZrCS}.  A UNF theoretical calculation with the code parameters constrained by the deduced data of the $^{93}$Zr($n$, $\gamma$)$^{94}$Zr reaction at $E_n$$>$3 MeV also agrees well with the directly measured cross sections.

\subsection{Measurement}
The experiment was carried out at the Tandem-accelerator located at the Japan Atomic Energy Agency (JAEA). The details of the experimental set-up can be found in \cite{YAN16}. An $^{18}$O beam with the energy of 117 MeV bombarded into the isotopically enriched zirconium target, which was made in the form of self-supporting metallic foil. The $^{90}$Zr target had a thickness of 300 $\mu$g/cm$^{2}$ and an isotopical enrichment of 99.4\%; while the $^{94}$Zr target had a thickness of 350 $\mu$g/cm$^{2}$ and an isotopical enrichment of 96.3\%. Downstream of the target, a silicon detector telescope $\Delta$$E-E$ \citep{NIS15} was used to identify the light ejectile particles, and two LaBr$_3$(Ce) detectors with the size of 4 inch in diameter and 5 inch in length were used for $\gamma$-ray detection \citep{MAK15}. A faraday cup was installed to collect the $^{18}$O beam current for normalization purpose. The $^{18}$O beam was kept to about 5 enA, the size of beam spot was less than 3 mm in diameter.

Beam time for each Zr target was about two days, and the accumulated number of $^{16}$O was roughly 5.8$\times$10$^5$ and 1.2$\times$10$^6$ for the $^{94}$Zr and $^{90}$Zr targets, respectively. The detected $\gamma$ ray events from $^{96}$Zr* and $^{92}$Zr* were about 5.9 $\times$ 10$^2$ and 1.9 $\times$ 10$^3$, respectively.

\subsection{Data Analysis}
\begin{figure}[ht!]
\figurenum{2}
\label{DEEplot}
\plotone{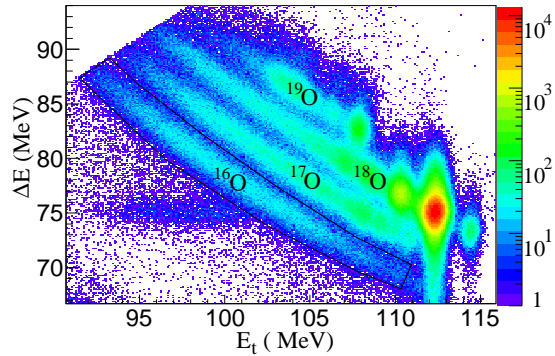}
\caption{Scatter plot of energy loss vs. total energy of the reaction products from $^{18}$O + $^{94}$Zr.}
\end{figure}

The recoil $^{16}$O was used to reconstruct the excitation energy $E_x$ of $^{92}$Zr$^*$ or $^{96}$Zr$^*$ by two-body kinematics, and the $\gamma$ rays were extracted in coincidence with the corresponding CN to obtain the $N_{\gamma(^{96}Zr^*)}(E_{n})$ and $N_{\gamma(^{92}Zr^*)}(E_{n})$ in Eq. \ref{csratio}.
The two-dimensional scatter plot of energy loss ($\Delta$$E$) versus total energy ($E_t$) was used to identify the recoil $^{16}$O. $E_t$ is the sum of energy loss in the $\Delta$$E$ detector and the residual energy in the silicon ring detector. As an example, the $\Delta$$E-E_t$ scatter plot corresponding to the second inner ring of the annular $E$ detector is shown in Fig. \ref{DEEplot}, with a cut to select $^{16}$O events from the ($^{18}$O,$^{16}$O) two-neutron transfer reaction. The energy resolution for $^{16}$O was about 1 MeV in full width at half maximum (FWHM), which was mainly due to the noise of silicon detectors and the kinematic uncertainty originated from the 1.2 degree resolution of each ring.

\begin{figure}[ht!]
\figurenum{3}
\label{ratio}
\plotone{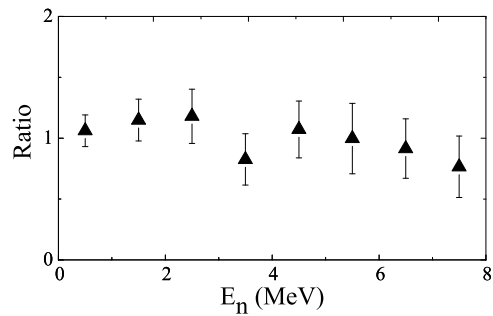}
\caption{The $\gamma$-decay probability ratio of the compound nuclei $^{96}Zr^*$ and $^{92}Zr^*$.}
\end{figure}

Since $^{92}$Zr and $^{96}$Zr are both even-even nuclei, the de-excitation of their high-lying resonance states is expected to proceed overwhelmingly through the first 2$^+$ state to the 0$^+$ ground state doorway transition. The energy of the transition is 1750 keV for $^{96}$Zr*, and 934 keV for $^{92}$Zr*. First, we deduced the net areas of these two $\gamma$ lines in a bin width of $E_n= 1000$ keV. The net areas were then normalized to the integrated $^{18}$O beam current, the target thickness, and the absolute detection efficiency of the LaBr$_3$ detectors, for each of the $^{94}$Zr($^{18}$O,$^{16}$O$\gamma$)$^{96}$Zr and $^{90}$Zr($^{18}$O,$^{16}$O$\gamma$)$^{92}$Zr run. The minor difference of solid angle due to the tiny difference of reaction kinematics for the $^{94}$Zr and $^{90}$Zr target runs was also taken into account. The absolute branching ratios of the 1750 keV and 934 keV $\gamma$ lines in $^{96}$Zr and $^{92}$Zr were taken as 80.0\% and 91.7\% respectively from the in-beam $^{96}$Zr(p, p$^{'}$$\gamma$) study \citep{MOL89} and the prompt $\gamma$-ray spectroscopy study \citep{NAK07} of the $^{91}$Zr($n$,$\gamma$)$^{92}$Zr reaction. After the corrections were made, the ratio of the $\gamma$ decay probabilities $N_{^{96}Zr^{*}\gamma}(E_{n})$/$N_{^{92}Zr^{*}\gamma}(E_{n})$ was obtained, as shown in Fig. \ref{ratio}. The energy resolution in the equivalent neutron energy $E_n$ is about 1 MeV, and altogether eight ratios were deduced within $E_n$=0-8 MeV.

\subsection{Experimental Cross Sections}

\begin{figure}[ht!]
\figurenum{4}
\label{95ZrCS}
\plotone{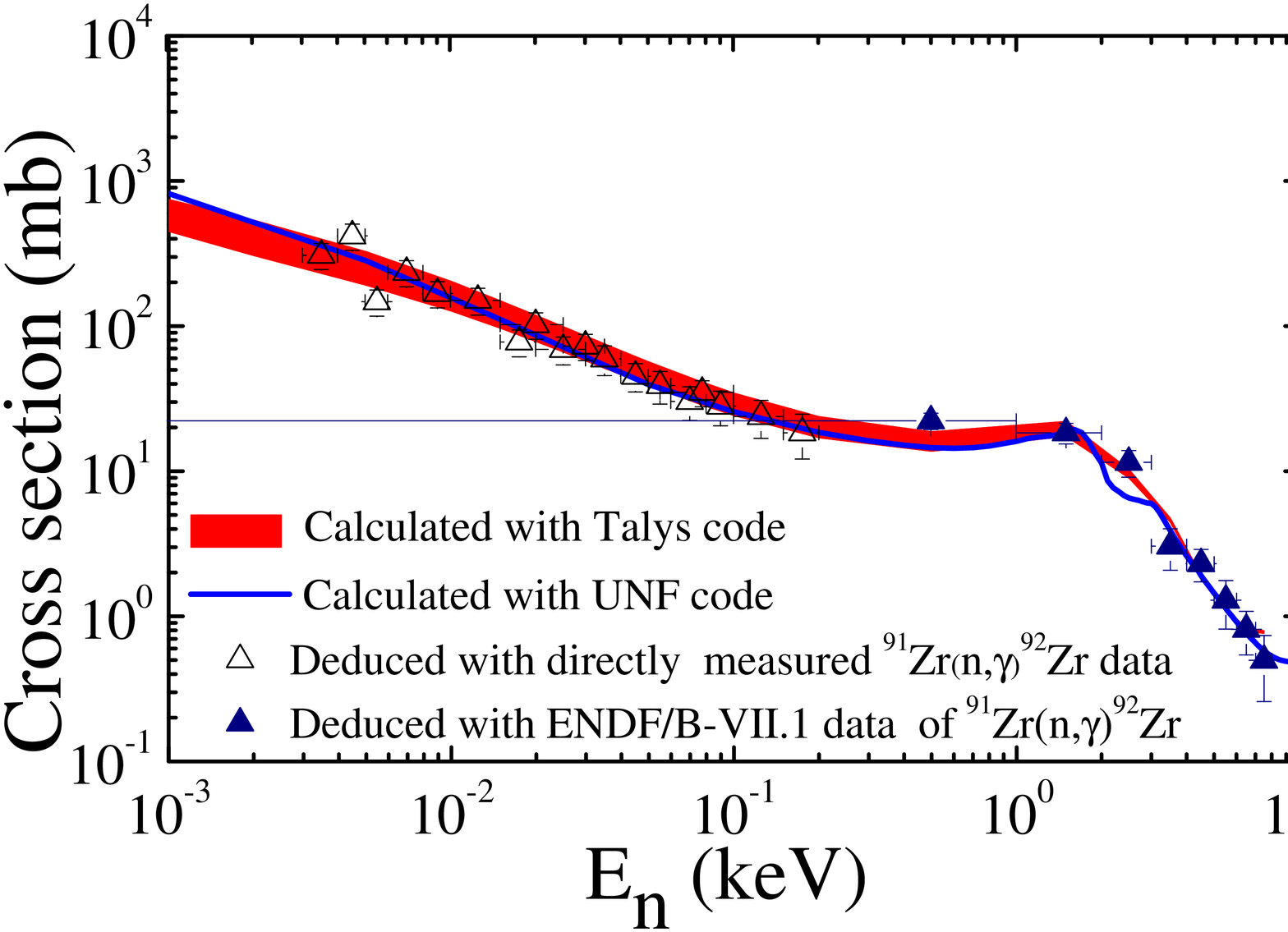}
\caption{Cross section of $^{95}$Zr($n$,$\gamma$)$^{96}$Zr.  The open triangles are deduced by multiplying our $\gamma$-decay probability ratio with the directly measured $^{91}$Zr($n$,$\gamma$)$^{92}$Zr cross section, the blue triangles are deduced by multiplying our $\gamma$-decay probability ratios with the ENDF data for the $^{91}$Zr($n$,$\gamma$)$^{92}$Zr cross section.}
\end{figure}

For nuclear astrophysics and other applications focusing on ($n$, $\gamma$) reactions, it is important to deduce the ($n$, $\gamma$) cross sections for $E_n$ $<$ 1 MeV. The excitation functions of $^{91}$Zr($n$,$\gamma$)$^{92}$Zr and $^{95}$Zr($n$,$\gamma$)$^{96}$Zr calculated by Hauser-Feshbach theory have a similar behavior at $E_n$ $<$ 1 MeV. The difference between the two ($n$,$\gamma$) cross sections can be approximated by a constant in this energy range, the $^{95}$Zr(n,$\gamma$)$^{96}$Zr cross sections can be determined by multiplying the average ratio of $1.06 \pm 0.21$ at $E_n$ $<$ 1 MeV to the directly measured $^{91}$Zr($n$,$\gamma$)$^{92}$Zr cross sections \citep{MUS77}. The uncertainty of the deduced $^{95}$Zr($n$,$\gamma$)$^{96}$Zr cross section includes the experimental uncertainty and an additional systematic uncertainty of 10-15\% which is estimated by UNF/TALYS \citep{KONING} code by a variation of the parameters of the calculation within a reasonable range. A further essential constraint of the parameter space was obtained from the analysis of the $^{96}$Zr($\gamma$,n)$^{95}$Zr data \citep{UTS10}(see next Section \ref{sec:theo}). The deduced cross sections are shown in Fig. \ref{95ZrCS} as open triangles.

According to \cite{CHI10}, the $\gamma$-decay probability ratio is relatively insensitive to the spin-parity distribution of CN at neutron energies $E_n$$>$3 MeV. The cross sections in high energy region could be used to constrain the parameters of the theoretical calculation. Since there are no experimental data at the neutron energies larger than 3 MeV, we used the ENDF/B-VII.1 cross section of the $^{91}$Zr($n$,$\gamma$)$^{92}$Zr reaction. The $^{95}$Zr($n$,$\gamma$)$^{96}$Zr reaction cross section was then deduced by the experimental ratio multiplied by the averaged value of the ENDF/B-VII.1 cross section for the $^{91}$Zr($n$,$\gamma$)$^{92}$Zr reaction. The deduced $^{95}$Zr($n$,$\gamma$)$^{96}$Zr cross sections are shown in Fig. \ref{95ZrCS} as blue triangles, the width of each energy bin is 1 MeV.

In the determination of ($n$,$\gamma$) cross sections with the SRM, the $\gamma$-decay probabilities are assumed to be independent from the spin-parities of CN in the Weisskopf-Ewing limit of the Hauser-Feshbach theory. However, theoretical study \citep{CHI10} shows that the $\gamma$-decay probability ratio is sensitive to the spin-parity distribution of the CN at low incident neutron energies and it is difficult to determine the ($n$,$\gamma$) cross sections with SRM because the surrogate reactions bring much more angular momentum to the CN compared to low-energy neutron. Similar to the present work, $^{91}$Zr($n$,$\gamma$)$^{92}$Zr and $^{93}$Zr($n$,$\gamma$)$^{94}$Zr were chosen to check the SRM with ($^{18}$O, $^{16}$O) two-neutron transfer reactions in the benchmark experiment described in Section 2.2 \citep{YAN16}. The $^{93}$Zr($n$,$\gamma$)$^{94}$Zr reaction cross section deduced by the SRM agrees well with the directly measured ones at $E_n$$<$1 MeV, which implies that the sensitivity of $\gamma$-decay probability ratio to the CN spin-parity distribution is partially reduced in SRM at low energies. Furthermore, the theoretical prediction of $^{93}$Zr($n$,$\gamma$)$^{94}$Zr reaction agrees with the directly measured value and with the SRM determined cross sections, and the agreement between theoretical calculations and SRM determined cross sections of $^{95}$Zr($n$,$\gamma$)$^{96}$Zr is supportive of our determination.

Finally, in Eq. \ref{cnratio}, the CN formation cross sections integrated over the detector solid angle ($\sigma^{CN}_{Zr^*}(E_{ex})$) of the two surrogate reactions, $^{94}$Zr($^{18}$O, $^{16}$O)$^{96}$Zr$^*$ and $^{90}$Zr($^{18}$O, $^{16}$O)$^{92}$Zr$^*$, are supposed to be the same in the formula. In the measurement we verified that the counts of CN $^{96}$Zr$^*$ and $^{92}$Zr$^*$, deduced by the outgoing particle $^{16}$O,  are almost same within 9\% at $E_n$$<$8 MeV after the corrections of beam current and target thickness.

\subsection{Theoretical Excitation Function} \label{sec:theo}

The calculation of the astrophysical reaction rate in the next section requires an excitation function of the $^{95}$Zr($n$,$\gamma$)$^{96}$Zr cross section over a sufficient energy range. In a first stage of the analysis we calculated this excitation function using the statistical model code UNF, and good agreement with the experimental results was found (see Fig. \ref{95ZrCS}). These UNF calculations were later extended using the code TALYS for an estimate of the uncertainties. For this purpose the complete parameter space of TALYS was investigated which includes variations of the gamma-ray strength function, the level density, and the nucleon optical model potential. Because of its negligible influence, the alpha optical model potential was taken from the new TALYS-V1.8 default in the present work. For details of the investigation of the TALYS parameter space, see a similar study on $\alpha$-induced cross sections on $^{64}$Zn \citep{MOHR17}.

For each combination of TALYS parameters $\chi^2$ values were calculated for the indirect $^{95}$Zr(n,$\gamma$)$^{96}$Zr data of the present work and for
the data of the reverse $^{96}$Zr($\gamma$,n)$^{95}$Zr reaction by \cite{UTS10}. These data for the reverse reaction were measured with high precision using monochromatic photons from Laser-Compton scattering. Later average cross sections \citep{CRA14,NAIK14} from bremsstrahlung are in rough agreement with the monochromatic photon data; because of their larger uncertainties, the bremsstrahlung data were not included in the present $\chi^2$-based adjustment of the TALYS parameters.

For the best fit an overall $\chi^2/F \approx 3$ was obtained which corresponds to an average deviation factor of 1.2 between the experimental data and the theoretical excitation functions. For the (n,$\gamma$) data $\chi^2/F \approx 1.3$ and a similar average deviation of about 1.3 is found; because of the larger uncertainty of the (n,$\gamma$) data, $\chi^2/F$ is smaller for a similar deviation. The ($\gamma$,n) data are described with $\chi^2/F \approx 3.5$ and a small average deviation of 1.1 only. The $\chi^2/F$ in the full parameter space of TALYS vary between its minimum value of 3 up to about 300, corresponding to a poor description of the data with an average deviation of a factor of three.

For an estimate of the uncertainty of the $^{95}$Zr(n,$\gamma$)$^{96}$Zr cross section all combinations of TALYS parameters were selected which describe the
(n,$\gamma$) and ($\gamma$,n) data with a reasonable $\chi^2/F \le (\chi^2/F)_{\rm{min}} + 1$; for a discussion of this choice, see \cite{MOHR17}. It is found that this choice essentially selects the gamma-ray strength function whereas the sensitivity of the calculated cross sections to the chosen level density and the chosen nucleon optical model potential remain minor. All reasonable $\chi^2/F$ are obtained from the simple Brink-Axel Lorentzian gamma-ray strength function \citep{KONING,BRI57,AXEL62} although recent work has shown that there are additional contributions to the gamma-ray strength function which may affect neutron-capture cross sections also close to magic numbers (e.g., \cite{CRES16}).

The results from all TALYS calculations with reasonable $\chi^2$ are shown in Fig. \ref{95ZrCS} as an error band in comparison with the data extracted by SRM for the $^{95}$Zr($n$,$\gamma$)$^{96}$Zr reaction. Because the TALYS parameters were adjusted to the experimental data, the excellent agreement is
not surprising. Interestingly, the median result of the TALYS calculations and the UNF calculation are very close within a few per cent. Thus, the first UNF
calculation was used for the calculation of the MACS in the following section. The uncertainty of the MACS of about 25 per cent is estimated from
the experimental uncertainty of the SRM data. A similar estimate for the uncertainty is obtained from the study of the TALYS parameter space.

\begin{figure}[ht]
\figurenum{5}
\label{96Zrgn}
\plotone{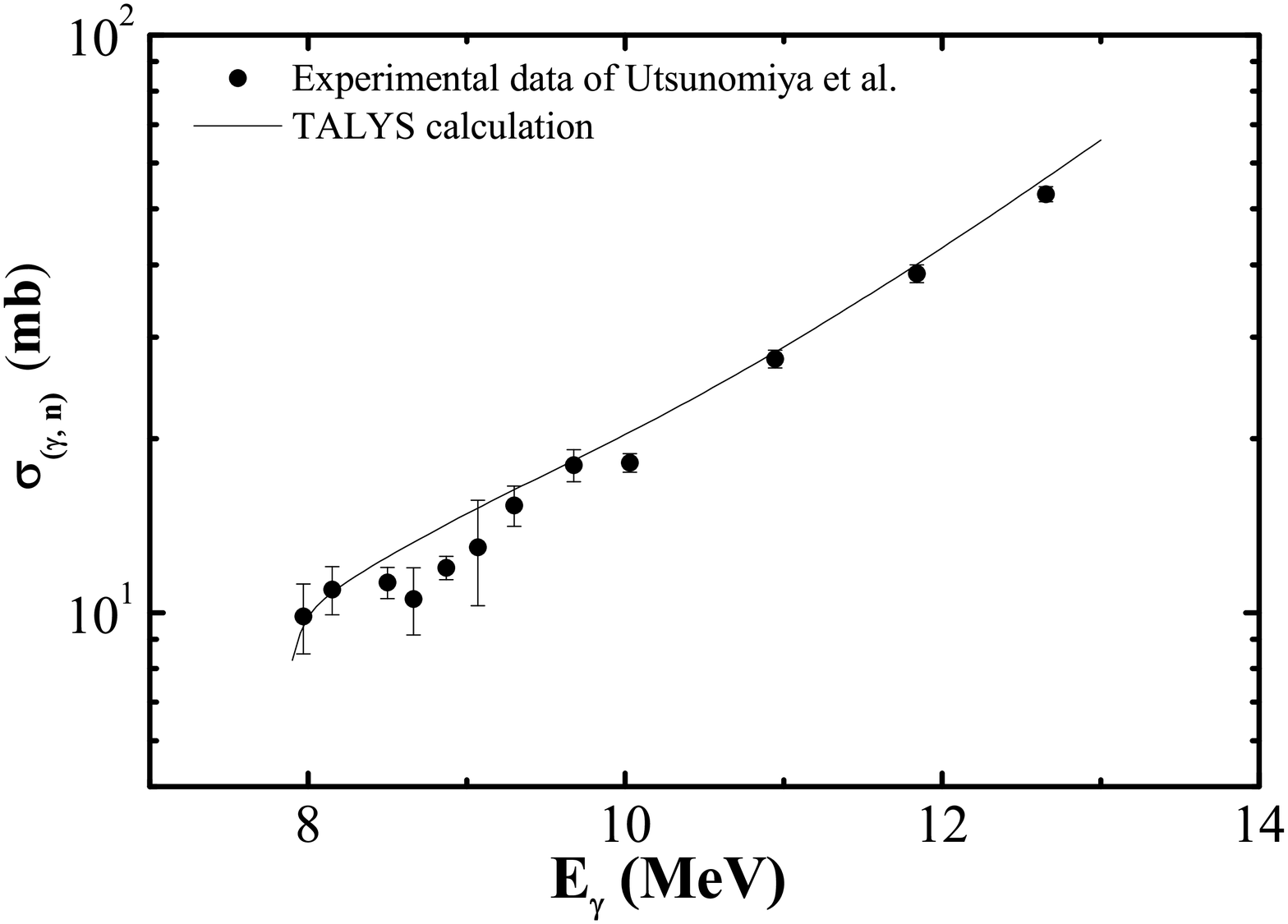}
\caption{Cross section of $^{96}$Zr($\gamma$,n)$^{95}$Zr. The experimental
  data of \cite{UTS10} are well described by the TALYS calculations of the
  present work. Note that for this presentation the average of the two
  analysis methods in \cite{UTS10} is shown.
}
\end{figure}

Finally, in Fig. \ref{96Zrgn} we compare the calculated $^{96}$Zr($\gamma$,n)$^{95}$Zr excitation function to the experimental results of \cite{UTS10}. There is good agreement over the full energy range. It is not surprising that all TALYS calculations with reasonable $\chi^2$ are practically identical because the same gamma-ray strength function is used, and the $^{96}$Zr($\gamma$,n)$^{95}$Zr cross section is practically only sensitive to the gamma-ray strength function but insensitive to other ingredients of the statistical model.

\subsection{Reaction Rate}
From the deduced cross sections of $^{95}$Zr($n$,$\gamma$)$^{96}$Zr reaction, we derived the Maxwellian-Averaged Cross Sections (MACS) shown in Table \ref{MACSs}. The MACS of this work is 66 $\pm$ 16 mb at $k$T = 30 keV. This value is in-between the \cite{BAO00} and \cite{TOU90},  but much larger than \cite{LUG14} and smaller than JENDL-3.2 and significantly smaller than ENDF/B-VII.1 and TENDL.

For completeness, we point out that the MACS and the astrophysical reaction rate are almost entirely defined by the laboratory cross section of the
$^{95}$Zr(n,$\gamma$)$^{96}$Zr reaction because the first excited state in $^{95}$Zr at an excitation energy of 954 keV has a negligible thermal
population at s-process temperatures. Thus, the stellar enhancement factor remains very close to unity, and the ground state contribution to the astrophysical rate is almost 100\% \citep{RAU11}. A suggested candidate for a very low-lying excited state in $^{95}$Zr at 23 keV has been definitely excluded by a high-resolution $^{94}$Zr(d,p)$^{95}$Zr experiment \citep{SONN03}.

\begin{table}
\caption{\label{MACSs}  Our derived MACS for $^{95}$Zr($n$,$\gamma$)$^{96}$Zr in mb. Also shown for comparison are the MACS by \cite{BAO00}, labeled as Bao, which is based on a semi-empirical analysis of the neighboring nuclei and the MACS from \cite{LUG14} (labelled  as Tagliente), based on the comparison of the trends shown by the odd and even numbered nuclei in the region as done by \cite{TOU90}, but on the basis of updated neutron capture cross sections  \cite{TAG08a,TAG08b,TAG10,TAG11a,TAG11b,TAG13}.}
\begin{tabular*}{80mm}{@{}@{}@{}@{}cccc}
\toprule
\raisebox{0ex}[4ex]{$\kappa$T (keV)} & Tagliente & This work & Bao \\
\hline
5  &255   & 251$\pm$46 & 296 \\
10 &107   & 153$\pm$31 & 185 \\
15 &66   & 113$\pm$24 & 136 \\
20 &46   & 90$\pm$20  & 109 \\
25 &35   & 76$\pm$18  & 91  \\
30 &28   & 66$\pm$16  & 79$\pm$12\\
40 &21   & 53$\pm$14  & 63  \\
50 &16   & 45$\pm$12   & 54  \\
60 &15   & 40$\pm$11   & 47  \\
80 &12   & 33$\pm$10  & 39  \\
100 &11   & 29$\pm$9   & 34  \\
\hline
\end{tabular*}
\end{table}
The corresponding reaction rate as a function of temperature $T_{9}$ (in unit of $10^{9}$ K) is fitted with the expression used in the astrophysical reaction rate library REACLIB:
 \begin{eqnarray}
  \label{Reaction reate}
&&N_A\langle\sigma\textit{v}\rangle=exp(17.9065+0.0025T_{9}^{-1}-0.6686T_{9}^{-1/3}\nonumber\\&&-2.5117T_{9}^{1/3}+1.5193T_{9}-0.3671T_{9}^{5/3}\nonumber\\&&-0.4058lnT_{9}).
\end{eqnarray}
The fitting errors are less than 2\% in the range from T$_{9}$ = 0.01 to T$_{9}$ = 2.

\section{Astrophysical implications}

As mentioned in the Introduction, during the $s$-process in AGB stars the activation of the branching point at
$^{95}$Zr controls the abundance of $^{96}$Zr. The branching factor, i.e., the probability that $^{95}$Zr
captures a neutron instead of $\beta$-decay, strongly depends on the maximum neutron density achieved in the star.
This is controlled mostly by the efficiency of the $^{22}$Ne($\alpha$,$n$)$^{25}$Mg neutron source, which
increases with increasing the stellar mass and decreasing the metallicity. It follows that the production of
$^{96}$Zr in AGB stars and the impact of the revised value of the neutron-capture cross section of $^{95}$Zr
depend on the stellar mass and metallicity considered. To illustrate this, in Table~\ref{rates}
we show the effect of the different neutron-capture cross section of $^{95}$Zr on the final abundance of
$^{96}$Zr for models of different masses (from 2 to 6 M$_{\odot}$) and two different metallicities, solar
(0.014) and roughly twice solar (0.03). A detailed description of the models can be found in \cite{KAR16}.
We run models using the $^{95}$Zr(n,$\gamma$)$^{96}$Zr values listed in Table \ref{MACSs}.

\begin{table}
\makebox[\textwidth][c]{}
\caption{\label{rates}
Variations in the final abundance of $^{96}$Zr (in units of $10^{-11}$) in selected stellar AGB models for different $^{95}$Zr(n,$\gamma$)$^{96}$Zr values. The percent variations with respect to the MACS of Tagliente are reported in brackets.}
  \begin{tabular*}{85mm}{@{}c@{}@{}@{}@{}ccccc}
  \toprule
 Mass & Metallicity & Tagliente & This work & Bao \\
 \hline
 2 M$_{\odot}$ & 0.014 & 0.95 & 1.18 (+25\%) & 1.28 (+35\%) \\
 3 M$_{\odot}$ & 0.014 & 3.57 & 6.39 (+79\%) & 7.37 (+106\%) \\
 6 M$_{\odot}$ & 0.014 & 1.09 & 1.24 (+14\%) & N/A$^a$  \\
 3 M$_{\odot}$ & 0.03 & 1.97 & 2.56 (+30\%) & 2.79 (+42\%) \\
 3.5 M$_{\odot}$ & 0.03 & 2.61 & 3.80 (+46\%) & 4.25 (+63\%) \\
 4 M$_{\odot}$ & 0.03 & 5.44 & 8.84 (+63\%) & 9.93 (+83\%) \\
\hline
\end{tabular*}
 $^a$This model was not run for time issues, as it takes almost 2 months. However, we expect a result close to
that obtained with the rate from this work.
\end{table}

In AGB stars of mass below roughly 3 M$_\odot$ the main neutron source is the $^{13}$C($\alpha$,$n$)$^{16}$O
reaction, which is activated at low temperatures $\sim$ 8 keV and produces low neutron densities, of the order
of $10^7$ cm$^{-3}$ \citep{STR97}. For neutron densities below $10^9$ cm$^{-3}$, the probability to
produce $^{96}$Zr is of the order of a few percent or less. It follows that in this mass range low abundances
of $^{96}$Zr are produced and the effect of changing the neutron-capture cross section of $^{95}$Zr is
relatively small, less than $\sim$30\%. As the stellar mass increases higher temperatures are reached and the
$^{22}$Ne($\alpha$,$n$)$^{25}$Mg reaction source is activated. This neutron source produces neutron densities
up to 10$^{13}$ cm$^{-3}$ in stars of mass around 6 M$_\odot$ \citep{RAA12}. For neutron densities above
10$^{11}$ cm$^{-3}$ the branching factor at $^{95}$Zr is roughly equal unity, which means that the main path
of neutron captures proceeds through $^{96}$Zr, instead of $^{95}$Mo. In this case the exact value of the
neutron-capture cross section of $^{95}$Zr does not play a major role in the production of $^{96}$Zr. Rather,
it is the value of the neutron-capture cross section of $^{96}$Zr itself that counts more, as the abundance of this isotope
reaches its equilibrium value. It follows that also in this case the effect on the abundance of $^{96}$Zr of
changing the neutron-capture cross section of $^{95}$Zr is small, 14\% for the 6 M$_{\odot}$ model.
The most interesting cases are the intermediate neutron densities values around 10$^{10}$ cm$^{-3}$ that can
be reached in AGB stars of masses between 3-5 M$_\odot$. For such value of the neutron density
the branching factor is roughly 0.5, i.e., there is 50\% probability that $^{95}$Zr will decay into $^{95}$Mo, and 50\% that is will capture a
neutron and produce $^{96}$Zr. In this regime the effect of changing the neutron-capture cross section of
$^{95}$Zr has a significant impact reaching differences up to a factor of two (Table~\ref{rates}).

In terms of the solar system abundance of $^{96}$Zr, \cite{TRA04} derived a contribution of roughly 40\% from
low-mass AGB stars and of roughly 40\% from massive AGB stars, using the neutron-capture cross
section for $^{95}$Zr from \cite{TOU90}. This is 50 mb at 30 KeV, comparable to the value presented here of
66 $\pm$ 16 mb. As illustrated above, the contribution from massive AGB stars does not change when changing
the rate, however, for typical low-mass (3 to 4 M$_\odot$) AGB models the contribution to the abundance of
$^{96}$Zr in the solar system would decrease by a factor up to 80\% when using
the lower (Tagliente) value of the rate by \cite{LUG14}.

Detailed constraints on the operation of the branching point at $^{95}$Zr
can be derived from the Zr isotopic ratios measured with high precision via mass
spectrometry in meteoritic stardust silicon carbide (SiC) grains. These grains originated in C-rich AGB stars
(i.e., with mass between roughly 1.5 and 4 M$_\odot$) of metallicity around solar and show the clear signature
of the $s$-process. However, to fully extract information from them on processes such as dust formation,
galactic chemical evolution, and the composition of the dust inventory present in the presolar nebula, we need
to accurately determine the mass and metallicity of their parent stars. This can be done using detailed
stellar nucleosynthesis models, where nuclear inputs such as the neutron-capture cross section of $^{95}$Zr
play a major role. \cite{LUG17} analysed in detail the comparison between the SiC data
and model for AGB stars of solar and twice-solar metallicity. They concluded that AGB stars of
metallicity twice-solar are good candidates for the origin of the grains, also based on the fact that the
$^{22}$Ne neutron source reaction in these stars is activated in such a way that the trend of the
$^{96}$Zr/$^{94}$Zr isotopic ratios can be well matched. This conclusion obviously depends on the choice of
the $^{95}$Zr($n$,$\gamma$)$^{96}$Zr rate.

\begin{figure}[ht!]
\centering
\figurenum{6}
\plotone{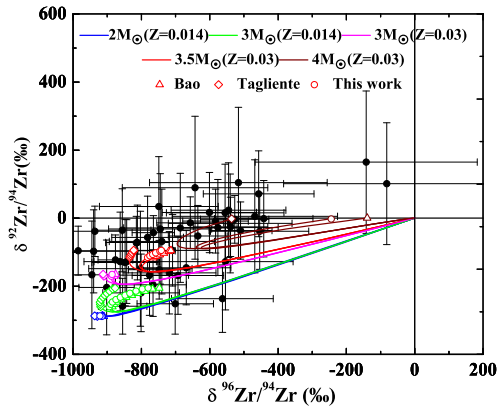}
\caption{The SiC grains data \citep[black circles with 2$\sigma$ error bars, for references see][]{LIU14}
are compared to the surface evolution of AGB stellar models different masses and metallicities from
\cite{KAR16}, using the standard choice of the parameter than controls the formation of the
$^{13}$C neutron source ($M_{\rm PMZ}$, see that paper for more details). For the comparison $\delta$-values are used,
which represent the permil variations with respect to the solar system ratio, corresponding to the initial
value used for the calculations. The straight, dashed lines represent the solar composition, with $\delta=0$ by definition.
Each coloured line represents the evolution of an AGB star of different initial mass and metallicity, as indicated
by the labels. Open symbols on the lines represent the phases when the envelope reaches C/O$>$1, the necessary condition for the formation of SiC, with each symbol corresponding to a different $^{95}$Zr($n$,$\gamma$)$^{96}$Zr rate, as indicated. \label{fig6}}
\end{figure}

In Fig. \ref{fig6}, we show the same selection of models from \cite{KAR16} presented in Table \ref{rates} and in comparison to the SiC grain data.

Because the value of the rate presented in this work is the same as the value from Bao, within the error bar, but it is significantly higher than the value by Tagliente, it results in $\delta$($^{96}$Zr/$^{94}$Zr) values ($\delta$($^{96}$Zr/$^{94}$Zr) = [($^{96}$Zr/$^{94}$Zr)$_{grain}$/($^{96}$Zr/$^{94}$Zr)$_{\odot}$ - 1] $\times$ 10$^3$) up to 400 permil higher than those computed with the rate by Tagliente. With the new rate the data can still be covered well, when considering the uncertainties in the neutron-capture cross section of $^{92}$Zr of 10\% at 2 sigma \citep{TAG10}, which would result in variation of $\pm$100 in the $\delta$($^{92}$Zr/$^{94}$Zr) values. Interestingly, using the new rate, the range of stellar mass that covers the data would have a maximum below 4 M$_\odot$, although this conclusion depends also on the uncertain rate of the $^{22}$Ne($\alpha$,$n$)$^{25}$Mg reaction \citep{MAS17}  and the adopted stellar model, as AGB models computed with different codes may vary somewhat in the temperature value that controls the activation of the $^{22}$Ne reaction.

\section{Summary and conclusion}
In summary, we have determined the $^{95}$Zr($n$,$\gamma$)$^{96}$Zr reaction cross section using the surrogate ratio method for the first time, and deduced a rate higher than that usually included in models of the s-process in AGB stars. The new rate was tested in the stellar models with stellar masses of 2 to 6 M$_\odot$ and metallicities of 0.014 and 0.03. We have seen that the largest effect when changing the rate are seen in models of masses around 3-4 M$_\odot$, where the $^{22}$Ne($\alpha$,$n$)$^{25}$Mg neutron source is mildly activated. We also concluded that using the new rate the masses of the parent stars of meteoritic stardust SiC grains should be somewhat lower than 4  M$_\odot$.

This demonstrates the potential of this method to derive better data for neutron capture cross sections of unstable nuclei relevant to branching points on the \textit{s}-process path. The method could be extended to other nuclei such as $^{85}$Kr, $^{107}$Pd and $^{181}$Hf with important implications on model predictions and the impact on the interpretation of observational constraints from spectroscopic to meteoritic data and the origin of specific isotopes such as $^{96}$Zr in the Solar System.

\vspace{10mm}

We thank JAEA Tandem accelerator facility staff for their help in the
experiment and the anonymous referee for his/her constructive suggestions. This work was supported by the National key Research and Development Program of China under Grant No. 2016YFA0400502, the National Natural Science Foundation of China under Grants No. 11375269, No. 11490560, No. 11321064, the National Basic Research 973 Program of China under Grant No. 2013CB834406, and the Hungarian Scientific Research Fund OTKA (K108459 and K120666). M.L. is a Momentum (Lend\"{u}let-2014 Programme) project leader of the Hungarian Academy of Sciences.This research was undertaken with the assistance of resources from the National Computational Infrastructure (NCI), which is supported by the Australian Government.

\software{UNF \citep{ZHA92,ZHA93,ZHA02}, TALYS \citep{KONING}.}





\end{CJK*}
\end{document}